\documentclass[a4paper]{jpconf}
\usepackage{graphicx}

\usepackage{amsmath}
\usepackage{amsfonts}
\usepackage{bbold}
\newcommand{\beq}{\begin{equation}}
\newcommand{\eeq}{\end{equation}}
\newcommand{\beqa}{\begin{eqnarray}}
\newcommand{\eeqa}{\end{eqnarray}}
\def\lsim{\raise0.3ex\hbox{$<$\kern-0.75em\raise-1.1ex\hbox{$\sim$}}}
\def\gsim{\raise0.3ex\hbox{$>$\kern-0.75em\raise-1.1ex\hbox{$\sim$}}}

\def\r{{\bf r}}

\def\0{{\bf 0}}

\def\kk{{\kappa}}
\def\pp{{\hat{p}}}

\begin{document}
\title{Heavy-flavour observables in relativistic nuclear collisions: theory overview}
\author{Andrea Beraudo}

\address{Istituto Nazionale di Fisica Nucleare - Sezione di Torino}

\ead{beraudo@to.infn.it}

\begin{abstract}
  Transport calculations represent the major tool to simulate the modifications induced by the presence of a hot-deconfined medium on the production of heavy-flavour particles in high-energy nuclear collisions. After a brief description of the approach  and of the major achievements in its phenomenological applications we discuss some recent developments. From the theory side we focus on the evaluation of transport coefficients and on recent formulations of the problem of heavy-flavour in-medium propagation in the language of open quantum systems. From a more phenomenological perspective we give an overview of the attempts to extend theoretical models to reproduce recent experimental data arising from event-by-event fluctuations (odd flow harmonics, event-shape-engineering) or from medium-modifications of hadronization ($D_s$ and $\Lambda_c$ production).     
  \end{abstract}

\section{Introduction}
Heavy-flavour particles play a peculiar role in probing the hot-deconfined matter produced in relativistic heavy-ion collisions. Soft observables (low-momentum light hadrons) provide information on the collective behaviour of the medium formed after the collision; they are nicely described by hydrodynamic calculations, assuming as a working hypothesis to deal with a system close to local thermal equilibrium. The suppression of the production of jets and high-$p_T$ particles tells us that a quite opaque medium is formed in the collisions: its description requires to model the energy-loss of high-energy partons in the hot plasma.
Heavy-flavour particles, arising with probability one from the hadronization of heavy quarks produced in initial hard events and having crossed the fireball during its whole evolution, require to employ a more general tool, allowing one to model their asymptotic approach to local thermal equilibrium with the medium: such a tool is represented by transport calculations, which we are going to briefly describe.
Actually, at high $p_T$, charm and beauty quarks play a different role, allowing one to study the mass and colour-charge dependence of parton energy-loss and jet-quenching: here we will not address this last important aspect.

\section{Transport calculations}
\begin{figure}[!ht]
\begin{center}
\includegraphics[clip,width=0.49\textwidth]{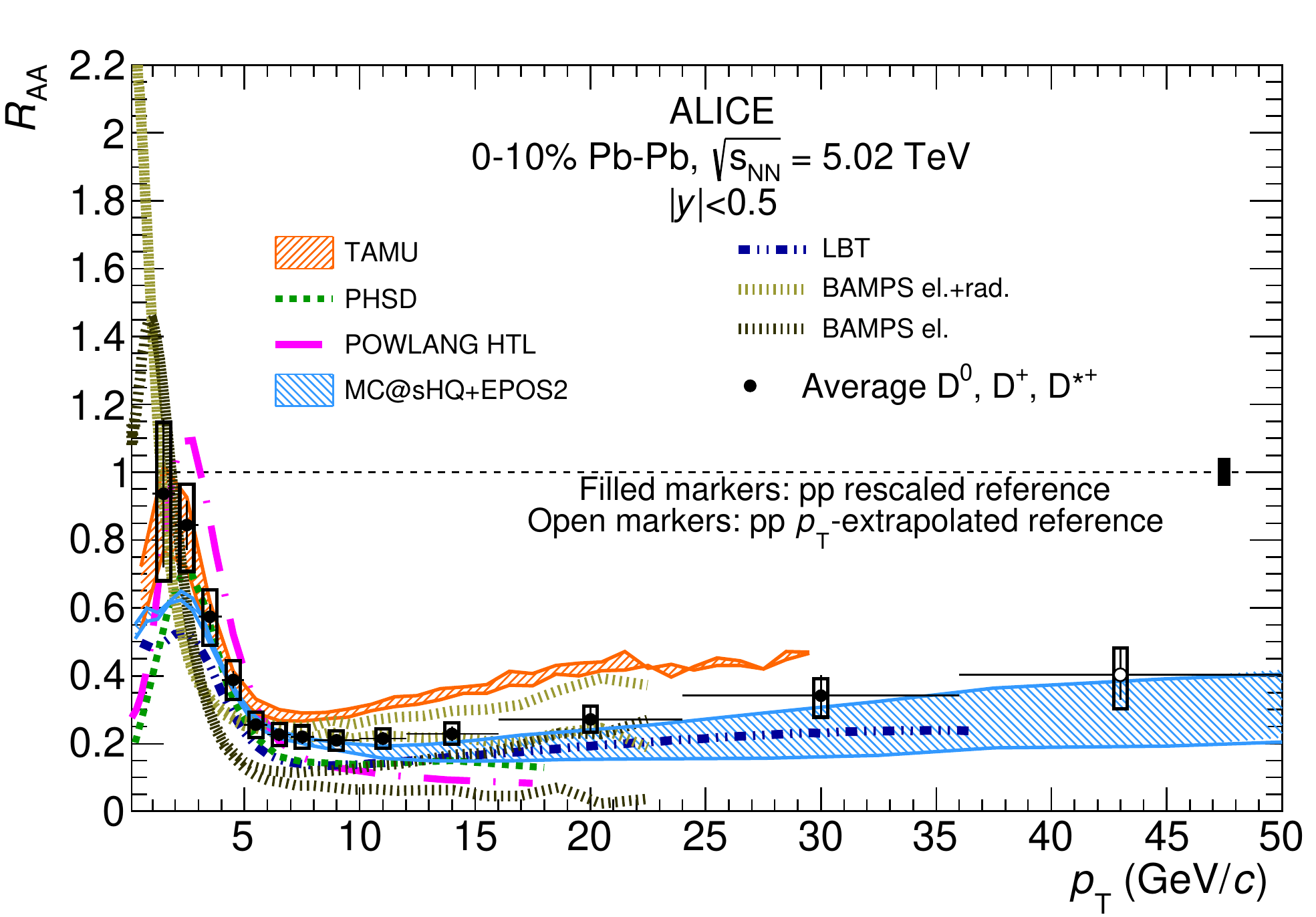}
\includegraphics[clip,width=0.49\textwidth]{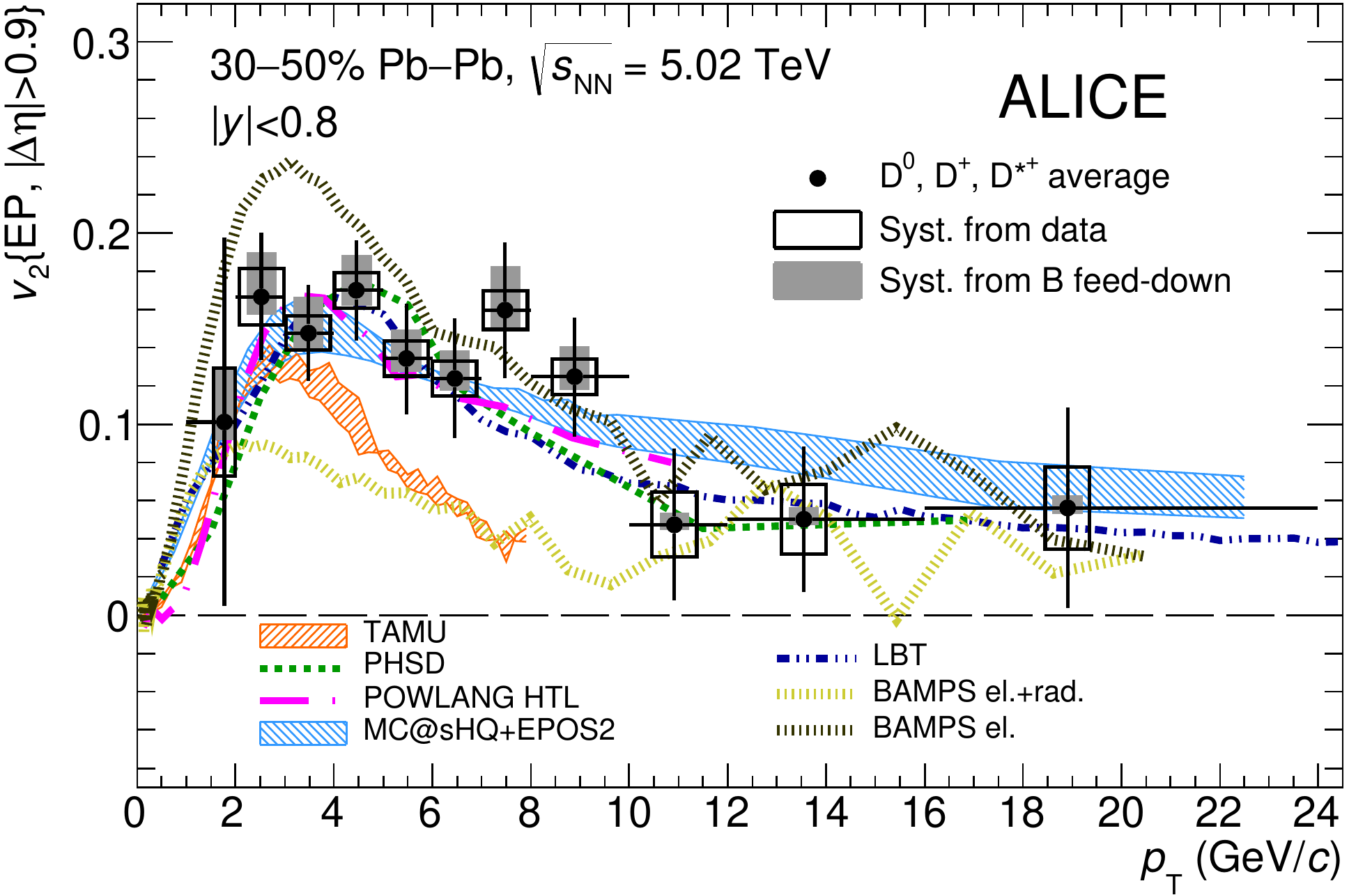}
\caption{Recent results by the ALICE collaboration for the average $D$-meson $R_{\rm AA}$ and $v_2$~\cite{Acharya:2018hre} compared to the predictions of various transport calculations.}\label{fig:data-vs-models} 
\end{center}
\end{figure}
The starting point of any transport calculation is the relativistic Boltzmann equation. Actually, in most numerical implementation, the latter is approximated as a more tractable Langevin equation, assuming that the heavy-quark interaction with the medium is dominated by multiple uncorrelated soft scatterings. One has then
\beq
{\Delta \vec{p}}/{\Delta t}=-{\eta_D(p)\vec{p}}+{\vec\xi(t)}.\label{eq:Langevin}
\eeq
Eq.~(\ref{eq:Langevin}) provides a recipe to update the heavy quark momentum in the time-step $\Delta t$ through the sum of a deterministic friction force and a random noise term specified by its temporal correlator
\beq
\langle\xi^i(\vec p_t)\xi^j(\vec p_{t'})\rangle\!=\!{b^{ij}(\vec p_t)}{\delta_{tt'}}/{\Delta t}\quad{\rm with}\quad{b^{ij}(\vec p)}\!\equiv\!{\kk_\|(p)}\pp^i\pp^j+{\kk_\perp(p)}(\delta^{ij}\!-\!\pp^i\pp^j).
\eeq
Following the heavy-quark dynamics in the medium requires then the knowledge of three transport coefficients representing the transverse/longitudinal ($\kappa_{\perp/\|}$) momentum broadening and the drag ($\eta_D$) received from the medium. Actually, the above coefficients are not independent, but are related by the Einstein fluctuation-dissipation relation, which ensures the asymptotic approach of the heavy quarks to thermal equilibrium.

Various transport calculations applied to heavy-flavour production in nuclear collisions can be found in the literature, essentially differing in the choice of transport coefficients to insert into Eq.~(\ref{eq:Langevin}). The challenge for the above models is to consistently reproduce various experimental observables, like the momentum and angular distributions of the produced heavy-flavour hadrons. A snapshot of the results of different models~\cite{Beraudo:2014boa,He:2014cla,Cao:2017hhk,Nahrgang:2013xaa} compared to ALICE data~\cite{Acharya:2018hre} is given in Fig.~\ref{fig:data-vs-models}. 

\begin{figure}[!ht]
\begin{center}
\includegraphics[clip,height=5.5cm]{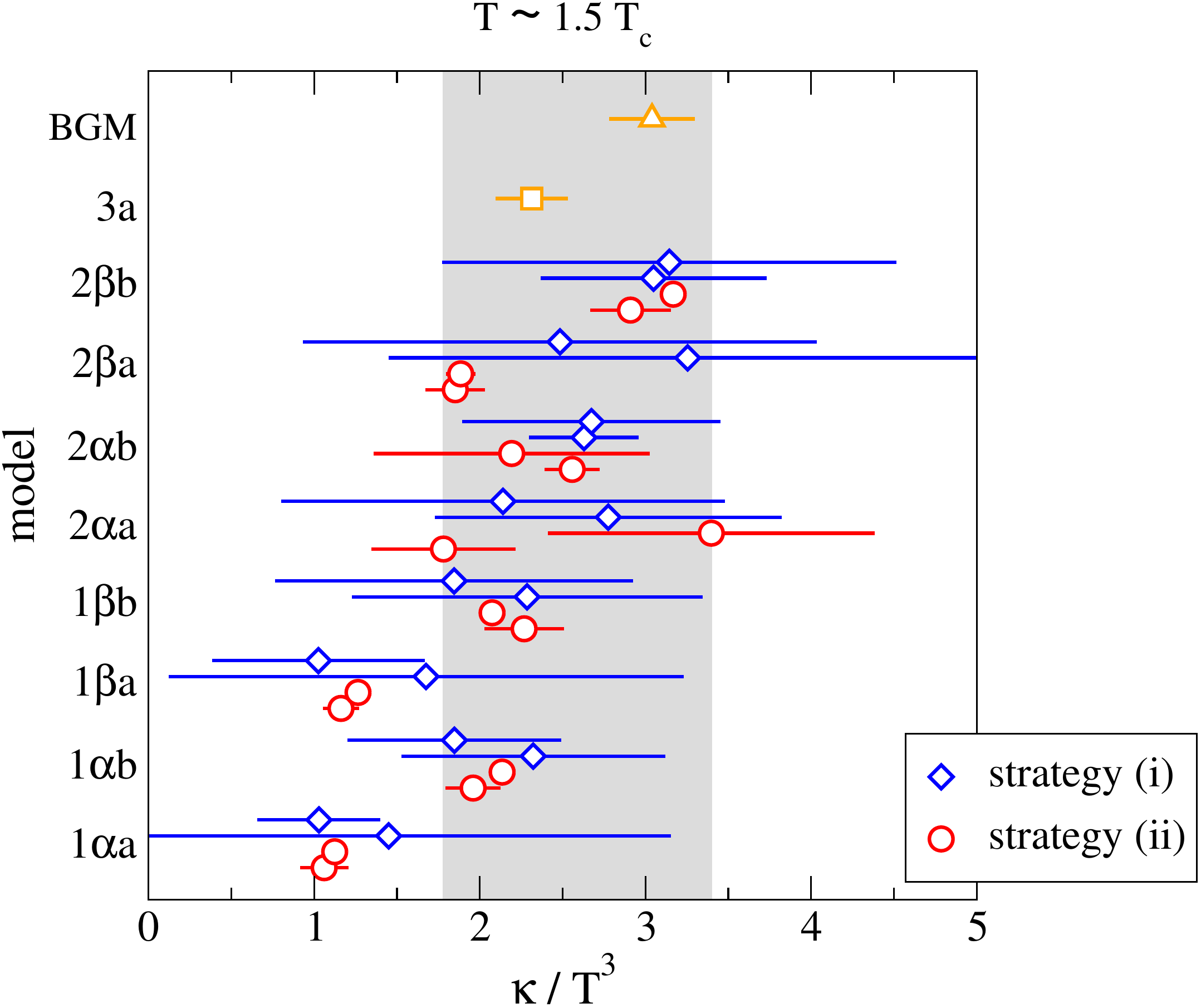}
\includegraphics[clip,height=5.5cm]{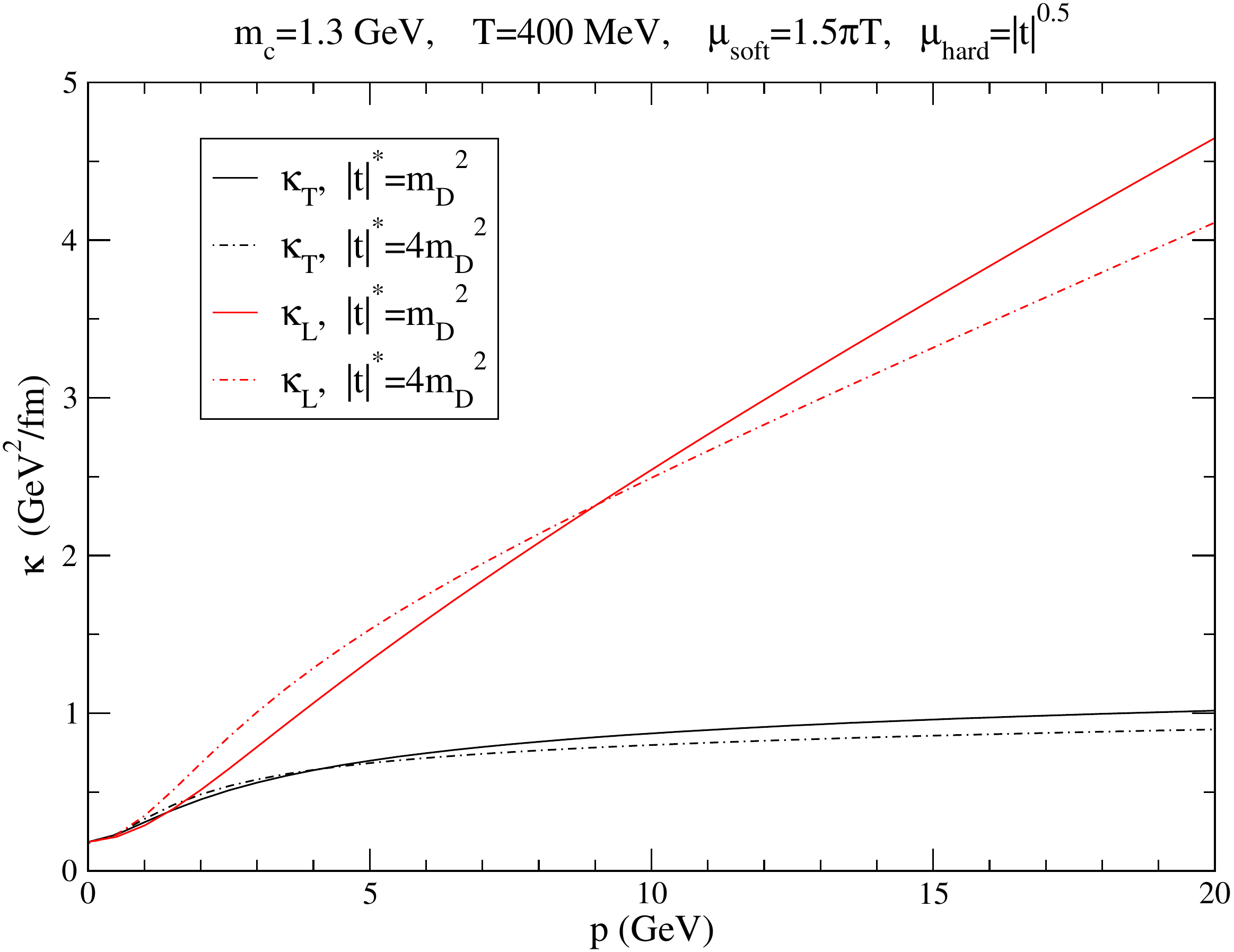}
\caption{Heavy-quark momentum diffusion coefficient $\kappa$ provided by different theoretical approaches: non perturbative lattice-QCD simulations (left panel) and weak-coupling thermal-field theory calculations (right panel). Notice the systematic uncertainty band of the lattice-QCD result~\cite{Francis:2015daa} and the different momentum broadening along the transverse and longitudinal directions predicted by the weak-coupling calculation~\cite{Alberico:2013bza}.}\label{fig:transp-coeff} 
\end{center}
\end{figure}
In principle the heavy-quark momentum-diffusion coefficient in the Quark-Gluon Plasma can be derived from the QCD Lagrangian.
A NLO weak-coupling calculation of $\kappa$ was performed in~\cite{CaronHuot:2008uh}, finding sizable positive corrections to the tree-level result for realistic values of the coupling. This suggests that a non-perturbative evaluation is in order. The latter can be provided by lattice-QCD simulations. The state-of-the-art results are presented in Ref.~\cite{Francis:2015daa} and displayed in the left panel of Fig.~\ref{fig:transp-coeff}. One notices the large systematic uncertainty of the estimate of $\kappa=(1.8-3.4)T^3$, arising from the extraction of real-time information from simulations performed in an Euclidean space-time. On top of the above problems, both the higher-order weak-coupling calculations and the lattice-QCD estimates suffer from a major limitation: they both deal with the case of a static infinitely-heavy quark at rest in the medium, while most of the current experimental data refer to charm quarks of relativistic momentum. In this regime one cannot neglect neither the momentum dependence of $\kappa$ nor the different behaviour of the transverse and longitudinal momentum broadening, as shown in the right panel of Fig.~\ref{fig:transp-coeff}, referring to a resummed weak-coupling calculation for charm quarks. Unfortunately, no higher-order calculation is so far available for the case of a finite momentum heavy quark.

\section{Insights from open-quantum-system theory}
\begin{figure}[!ht]
\begin{center}
\includegraphics[clip,height=5.5cm]{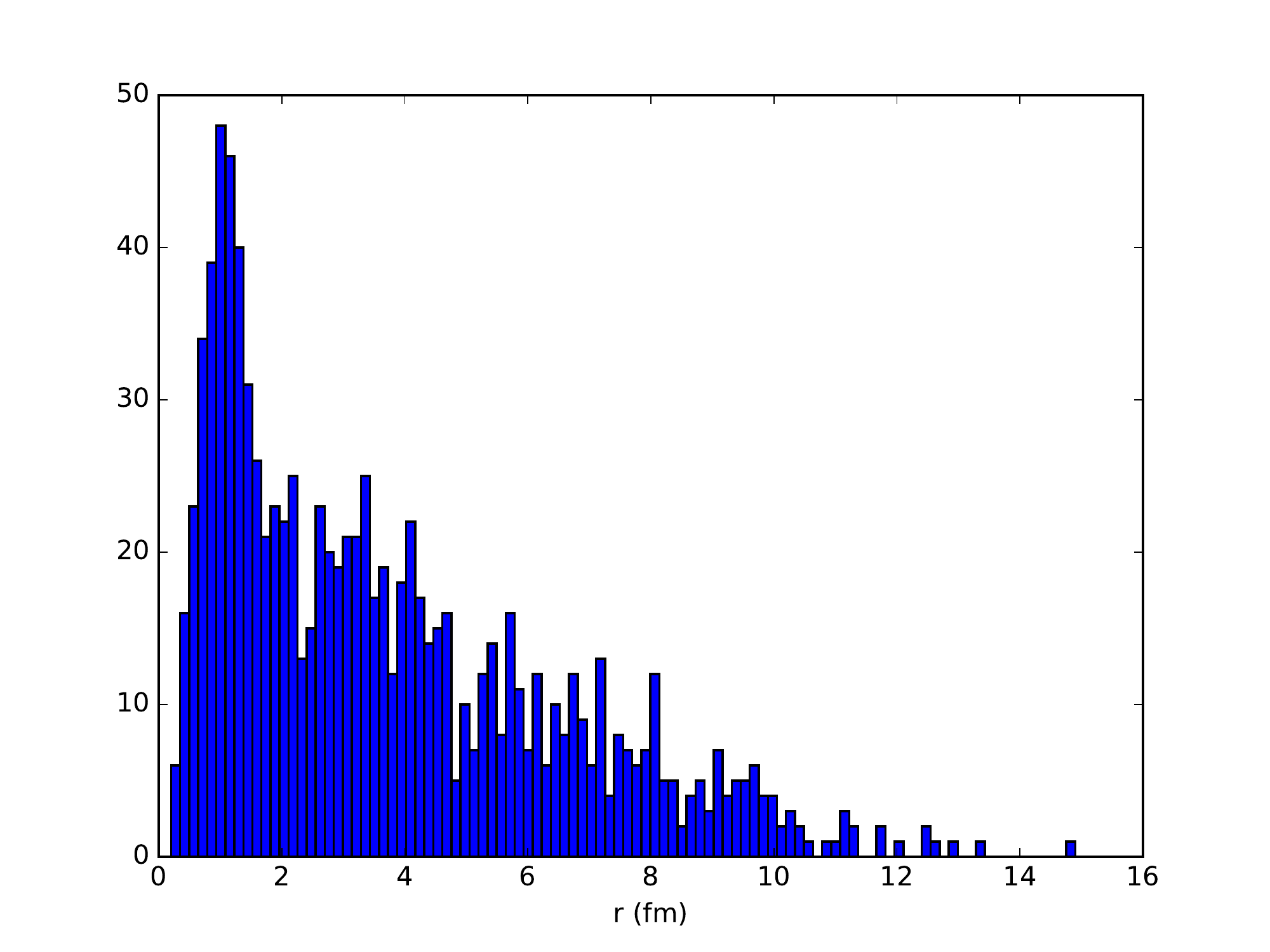}
\includegraphics[clip,height=4.4cm]{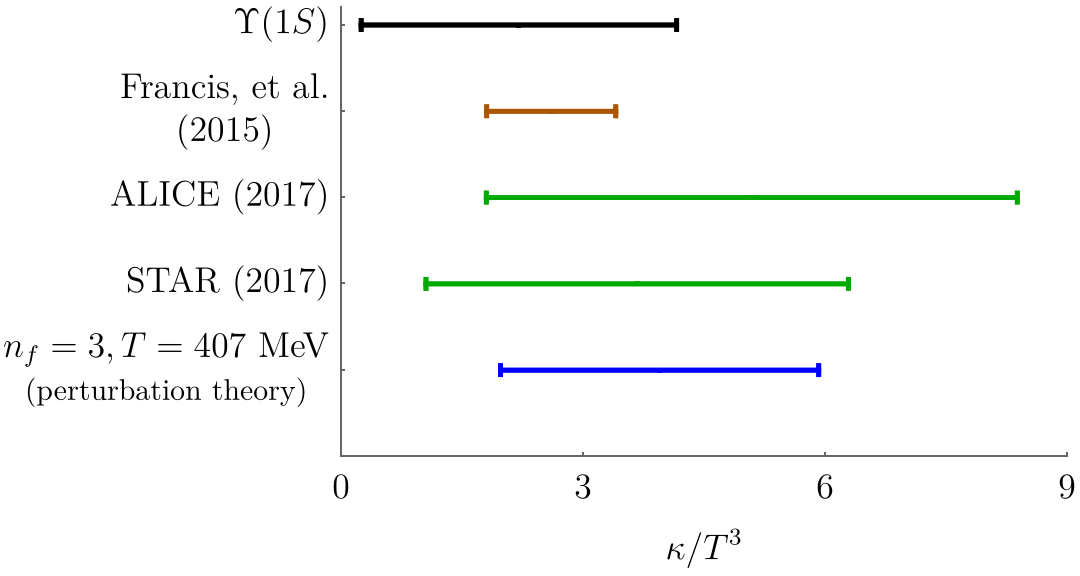}
\caption{Left panel: distribution of the relative distances for a sample of $Q\overline Q$ pairs, prepared in a 1S state, after propagating in a Quark-Gluon Plasma~\cite{Blaizot:2018oev}. Right panel: the momentum diffusion coefficient $\kappa$ extracted from the in-medium quarkonium width~\cite{Brambilla:2019tpt} compared to the results of other theoretical approaches and experimental estimates.}\label{fig:OQS} 
\end{center}
\end{figure}
In the last few years several authors~\cite{Blaizot:2018oev,Brambilla:2019tpt,Akamatsu:2014qsa,Yao:2018nmy} realized that heavy quarks (and quarkonia) in the medium can be conveniently treated as an open quantum system (the sample of Q's and $\overline Q$'s) coupled to an external environment (the hot plasma of gluons and light quarks). The full Hamiltonian reads then
\beq
    H=H_S\otimes\mathbb{1}_E+\mathbb{1}_S\otimes H_E+H_I.
    \eeq
    The general idea is that the full density matrix undergoes a unitary evolution
    \beq
     \frac{d\rho(t)}{dt}=-i[H_I(t),\rho(t)],   
     \eeq
     but this is no longer the case for the reduced density matrix of the system $\rho_S(t)\!\equiv\!{\rm Tr_E}\left(\rho(t)\right)$, obtained taking the trace over all possible states of the environment. The latter undergoes a non-unitary evolution described by the Lindblad equation.
     The above formulation of the problem allows a rigorous derivation of the Boltzmann equation, a consistent description of the evolution of heavy quarks and quarkonia and an alternative strategy to get a first-principle estimate of the heavy-quark transport coefficients.
     
     Performing a semi-classical approximation one can for instance derive a non-relativistic Langevin-like equation~\cite{Blaizot:2018oev}
    \begin{equation}
M\ddot{\bf r}_a=-C_F\gamma\dot{\bf r}_a+{\bf\Xi}_a(t)+\sum_{b\neq a}^{N_Q}{\bf \Theta}_{ab}(\r_{ab})+\sum_{\hat{b}}^{N_Q}{\bf\Theta}_{a\hat{b}}(\r_{a\hat{b}},t)\,,\label{eq:semiclass}
    \end{equation}
    where, besides the noise term due to the random collisions with the light degrees of freedom, one gets two additional terms due to the interaction with the other heavy $Q$'s and $\overline Q$'s present in the system. In the left panel of Fig.~\ref{fig:OQS} we display the relative distance distribution for a sample of $Q\overline Q$ pairs, initially produced in the 1S state, after evolving for a few fm/c in hot QGP according to Eq.~(\ref{eq:semiclass}).

    In Ref.~\cite{Brambilla:2019tpt} the authors showed that it is also possible to get an alternative estimate for the momentum diffusion coefficient $\kappa$ in terms of the Bohr's radius an the in-medium width of the quarkonium ground state, starting from the equation
    \beq
{\Gamma(1S)}=-2\langle{\rm Im(-i\Sigma_s)}\rangle={3a_0^2\kappa},\quad{\rm with}\quad    a_0\equiv\frac{2}{MC_F\alpha_s(1/a_0)}.
\eeq
The corresponding result for $\kappa$ is shown in the right panel of Fig.~\ref{fig:OQS}, together with the findings of independent estimates. The large systematic theoretical uncertainty $0.24\lsim \kappa/T^3\lsim 4.2$ arises from the necessity of extracting $\Gamma(1S)$ from a lattice-QCD in-medium spectral function.

\section{Recent developments: fluctuations, event-shape-engineering and directed flow}
\begin{figure}[!ht]
\begin{center}
\includegraphics[clip,height=6cm]{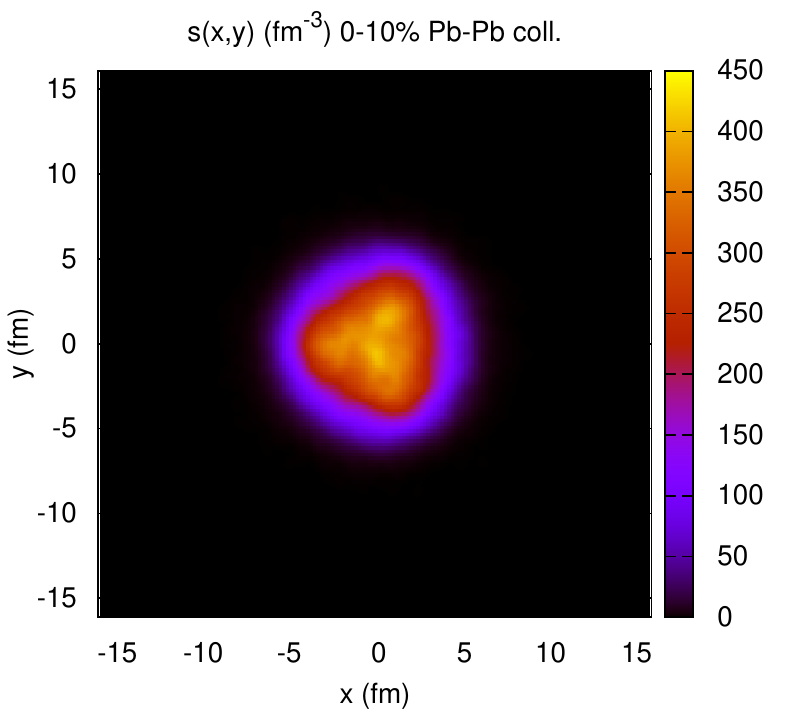}
\includegraphics[clip,height=6cm]{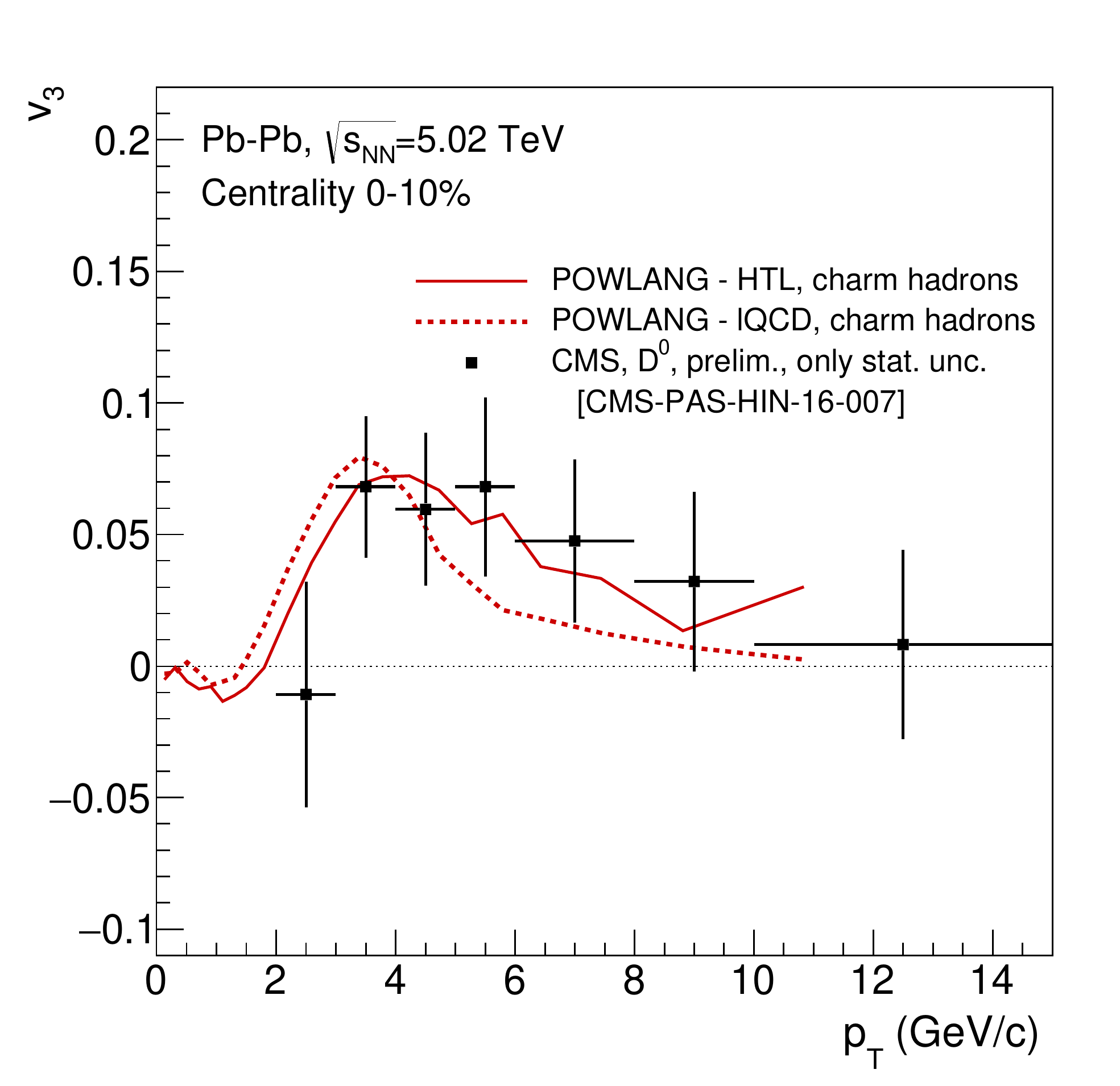}
\caption{Initial condition for a central Pb-Pb collision at the LHC displaying a triangular eccentricity (left panel) and the resulting $v_3$ coefficient (right panel) of the azimuthal distribution of $D$-mesons for different choices of the transport coefficients~\cite{Beraudo:2017gxw}.}\label{fig:e3-vs-v3} 
\end{center}
\end{figure}
The finite impact parameter of a nucleus-nucleus collision leads, on average, to an elliptic deformation of the produced fireball. Pressure gradients map this initial geometric asymmetry into a final momentum anisotropy of the particles decoupling from the medium, giving rise to the elliptic flow $v_2$ shown for instance in the right panel of Fig.~\ref{fig:data-vs-models}. However, event-by-event fluctuations (e.g. in the nucleon positions) can give rise to more complicated initial geometries, quantified by higher order eccentricity coefficients
\beq
\epsilon_{m}=\frac{\sqrt{\{r_\perp^2\cos(m\phi)\}^2+\{r_\perp^2\sin(m\phi)\}^2}}{\{r_\perp^2\}}
\eeq
which lead to higher harmonics in the final hadron distributions $v_M\equiv\langle\cos[m(\phi-\Psi_m)]\rangle$.
In Fig.~\ref{fig:e3-vs-v3} we show the result of a one-shot hydro+transport simulation starting from an average initial condition with a triangular deformation referring to the 0-10\% most central Pb-Pb collisions at $\sqrt{s_{\rm NN}}\!=\!5.02$ TeV. The final $D$-meson angular distribution is then characterized by a non-vanishing triangular flow $v_3$, as shown in the right panel of the figure where we compare the results of our transport simulations~\cite{Beraudo:2017gxw} to CMS data~\cite{Sirunyan:2017plt}.

\begin{figure}[!ht]
\begin{center}
\includegraphics[clip,height=5.2cm]{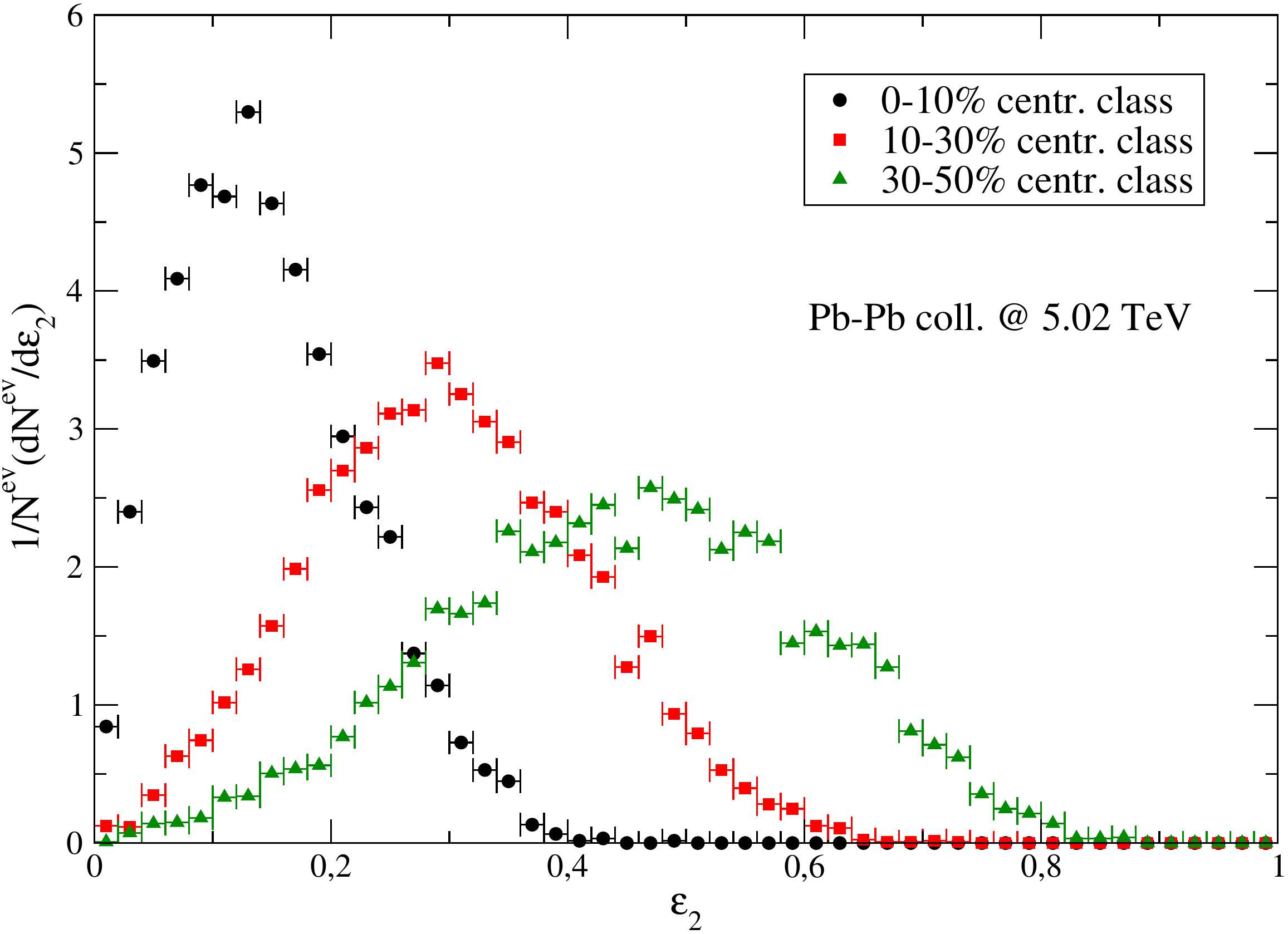}
\includegraphics[clip,height=5.2cm]{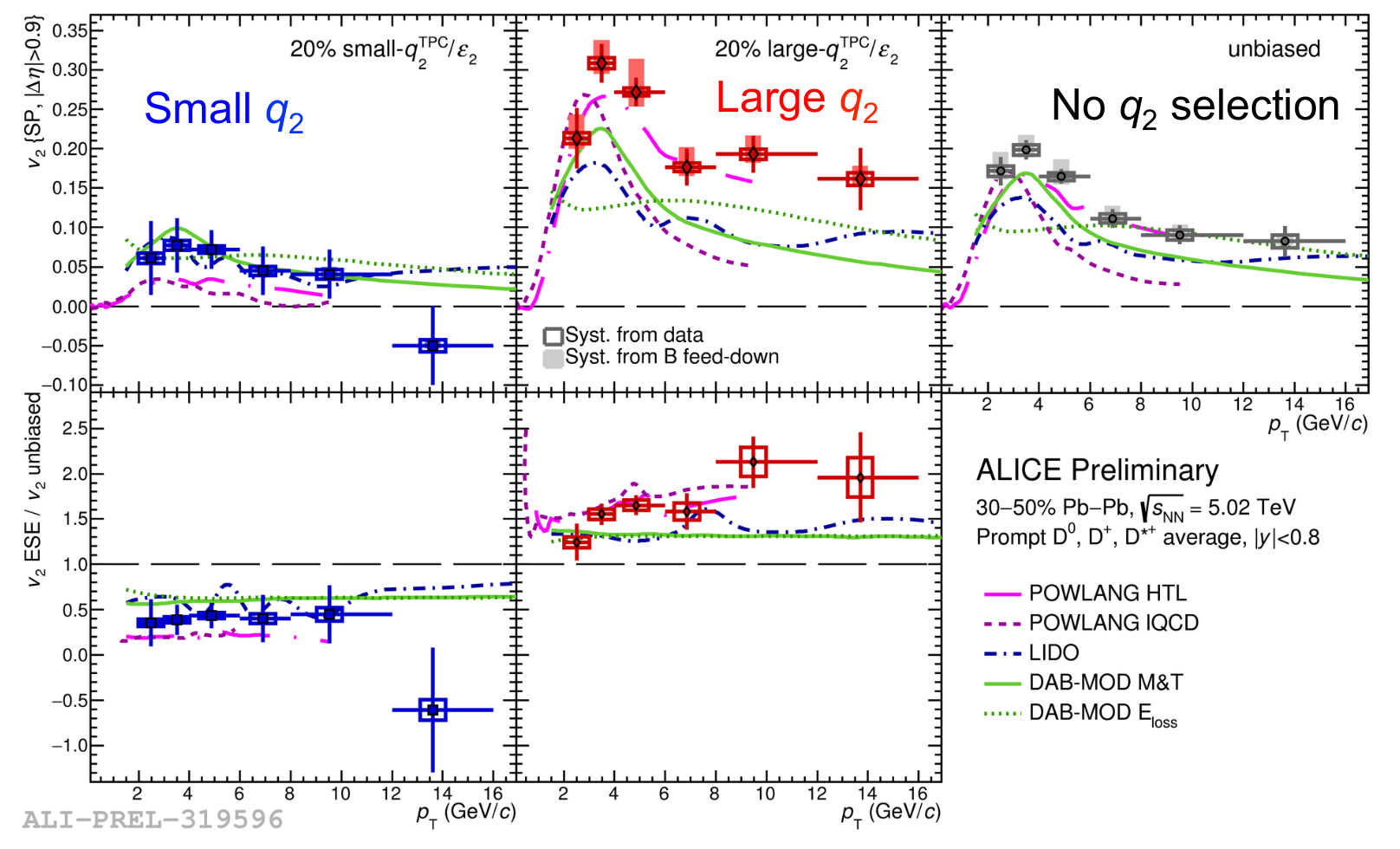}
\caption{Left panel: event-by-event initial eccentricity distribution in Pb-Pb collisions at the LHC from a Glauber-MC model~\cite{Beraudo:2018tpr}. Right panel: $D$-meson elliptic flow selecting events, within a given centrality class, with high/low eccentricity. ALICE data~\cite{Acharya:2018bxo} are compared to the predictions of various transport models.}\label{fig:ese} 
\end{center}
\end{figure}
\begin{figure}[!ht]
\begin{center}
\includegraphics[clip,width=0.9\textwidth]{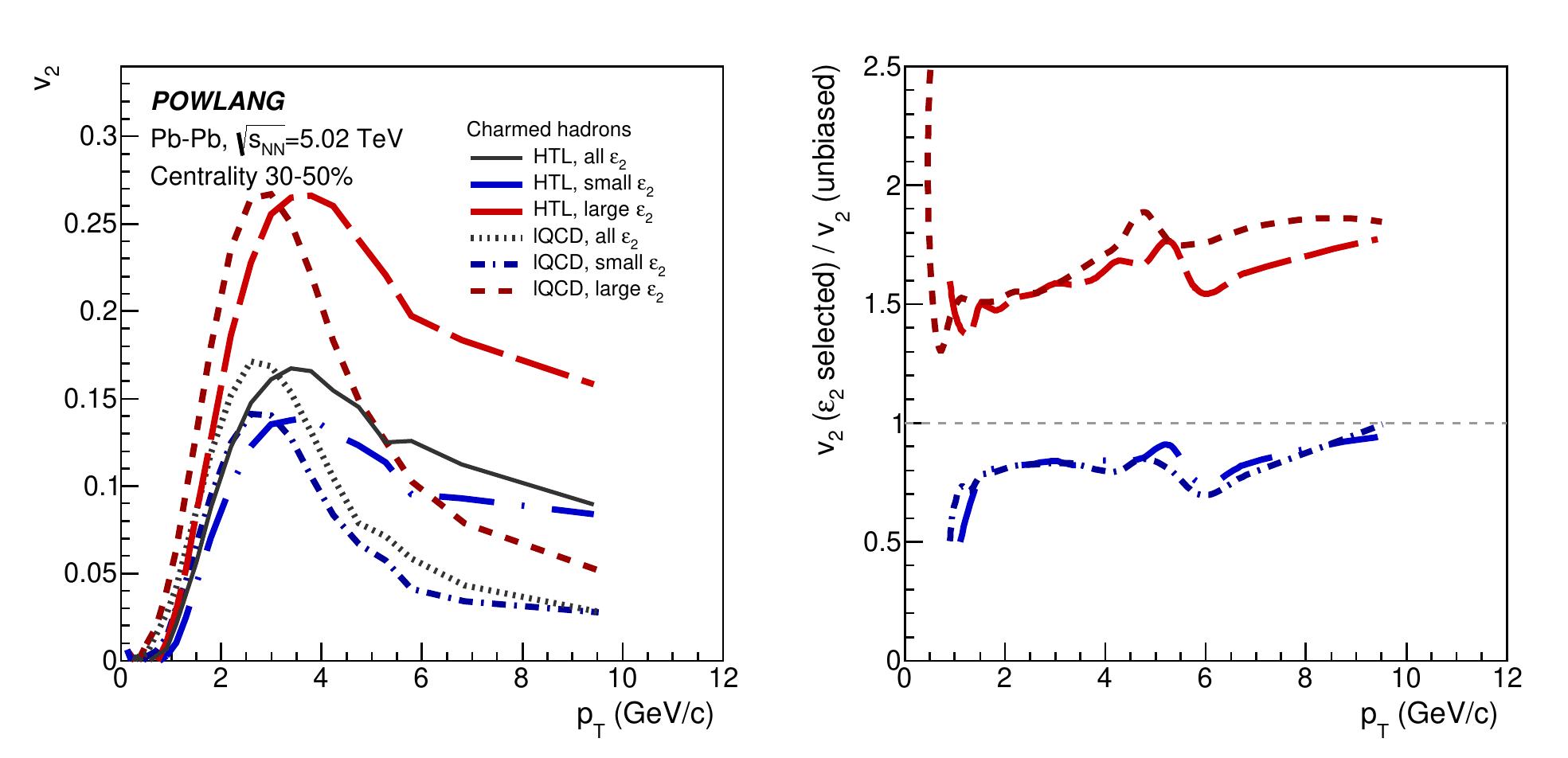}
\caption{$D$-meson elliptic flow in the 30-50\% centrality class for events with high/low initial eccentricity compared to an unbiased selection of events~\cite{Beraudo:2018tpr}. The ratio to the unbiased result (right panel) turns out to be independent of the choice of the transport coefficients.}\label{fig:ese2} 
\end{center}
\end{figure}
Due to event-by-event fluctuations, events belonging the same centrality class -- usually identified by some estimator like the number of binary nucleon-nucleon collisions or the multiplicity of produced particles -- can be characterized by quite different initial eccentricities, as shown in the left panel of Fig.~\ref{fig:ese}. It is then of interest to study, for a given centrality, the elliptic (or triangular) flow of the subsample of events of highest/lowest eccentricity, comparing the result to the unbiased case. This technique, known as \emph{event-shape-engineering}, was first introduced for light hadrons~\cite{Adam:2015eta} and later applied also to study of the flow of $D$-mesons~\cite{Acharya:2018bxo}. Results obtained by the ALICE collaboration, compared to various transport calculations, are displayed in the right panel of Fig.~\ref{fig:ese}. In Fig.~\ref{fig:ese2} we show the results of the transport model of Ref.~\cite{Beraudo:2018tpr} for the $D$-meson elliptic flow in Pb-Pb collisions, in the 30-50\% centrality class, for the 0-20\% highest and 0-60\% lowest-eccentricity subsamples. Notice how the ratio to the unbiased result does not depend on the choice of the transport coefficients, reflecting only the initial geometry.

\begin{figure}[!ht]
\begin{center}
\includegraphics[clip,height=4.8cm]{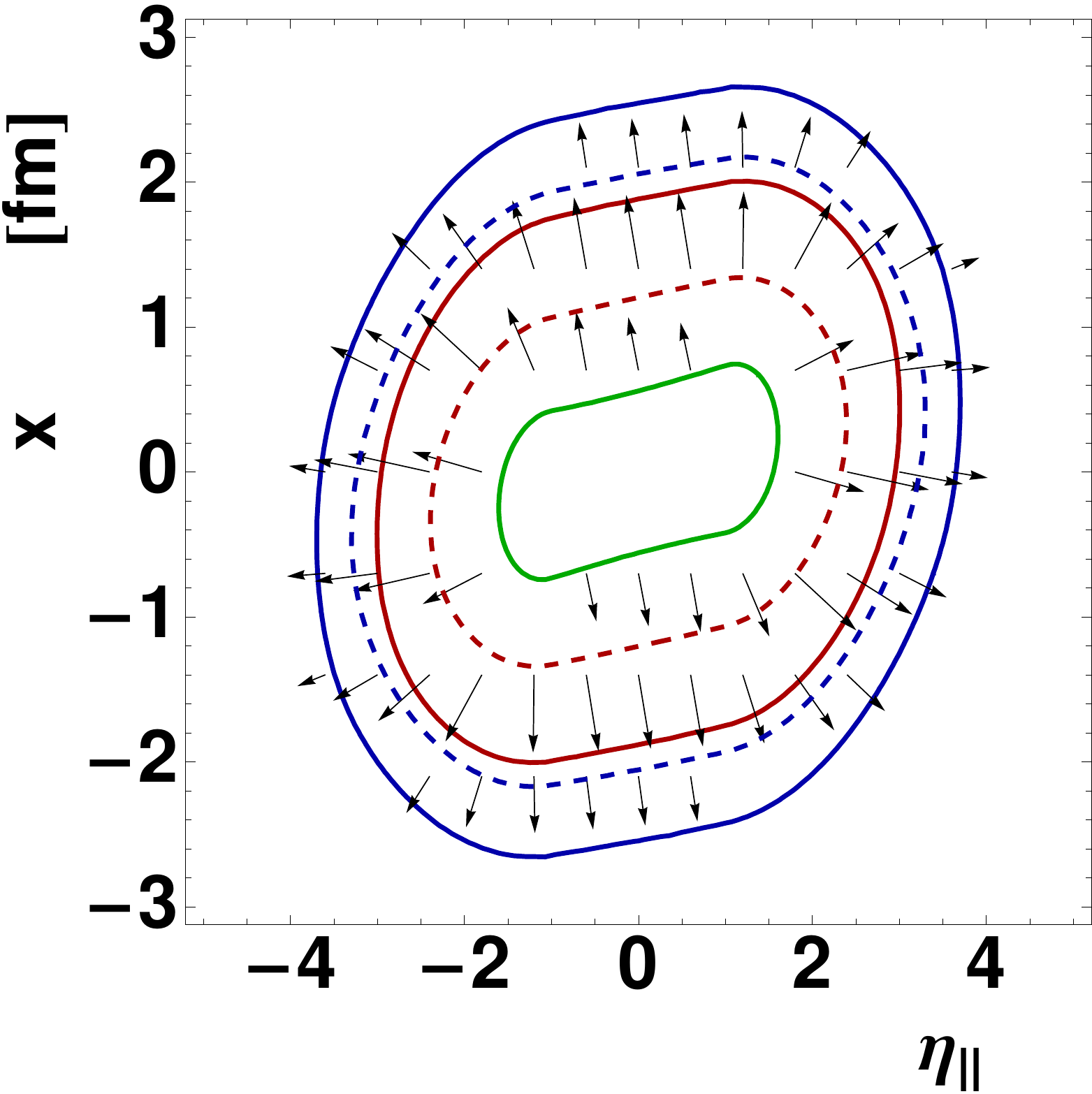}
\includegraphics[clip,height=4.8cm]{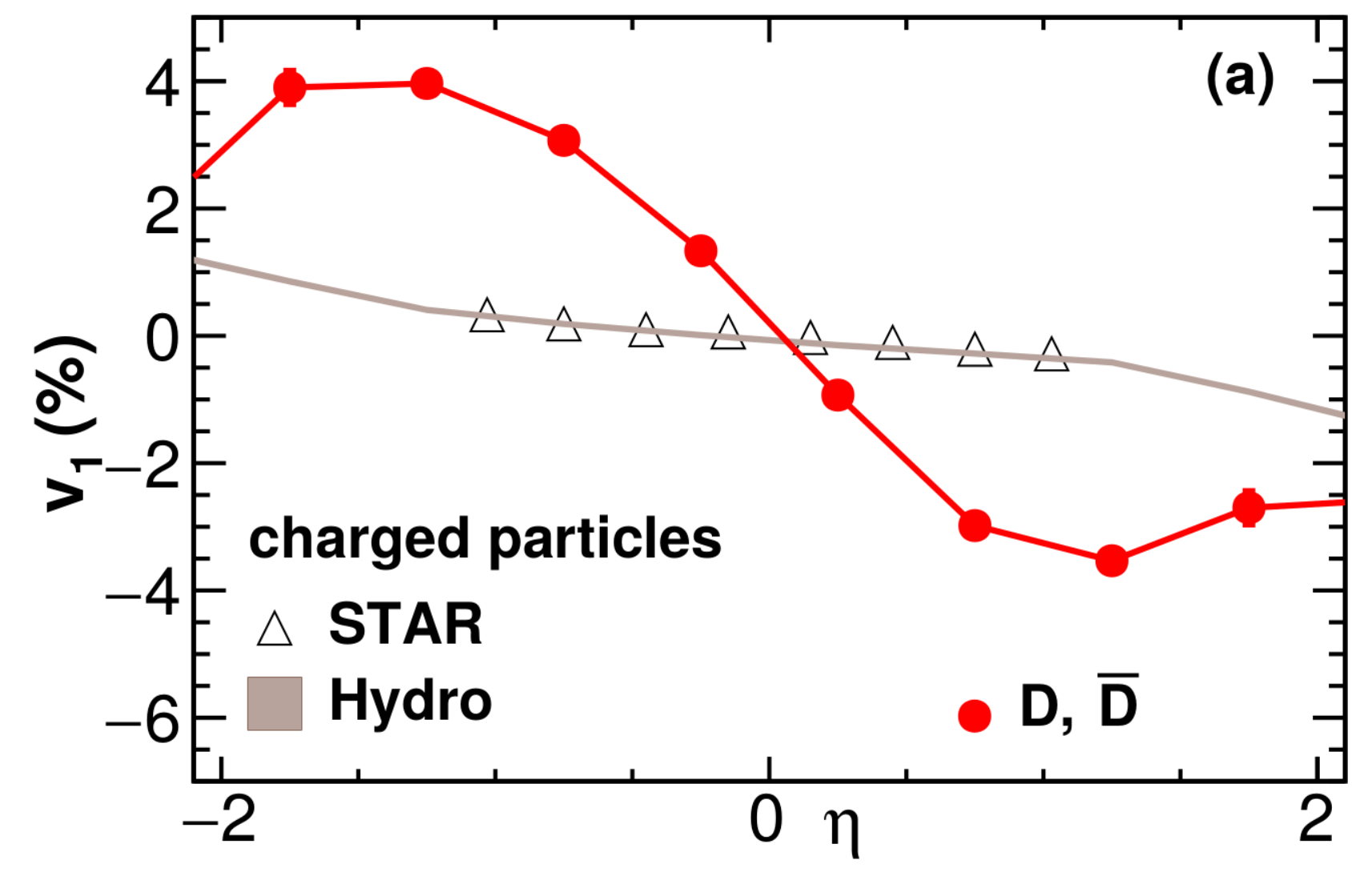}
\caption{Left panel: tilted initial condition in the $(x-\eta)$ plane (the space-time rapidity $\eta$ plays the role of a longitudinal coordinate, along the beam axis) for non-central nucleus-nucleus collisions~\cite{Bozek:2010bi}. Right panel: the directed flow of $D$-mesons as a function of pseudorapidity arising from a (3+1)D transport calculation~\cite{Chatterjee:2017ahy}.}\label{fig:v1} 
\end{center}
\end{figure}
Recently, a strong interest is growing also for the study of the directed flow $v_1\!\equiv\!\langle\cos(\phi\!-\!\Psi_{\rm RP})\rangle$. Since participant nucleons of the colliding nuclei tend to deposit more energy along their direction of motion, in non-central collisions the fireball is characterized by an initial tilted geometry (see left panel of Fig.~\ref{fig:v1}) and by a sizable orbital angular momentum (of order 1000$\hbar$) and vorticity~\cite{Becattini:2015ska}. Experimentally, this can give rise to a negative/positive directed flow $v_1$ of charged hadrons at forward/backward rapidity and possibly to other effects like the polarization of $\Lambda$ hyperons~\cite{STAR:2017ckg}. Interestingly, one expect a stronger $v_1$ signal (quite small for light hadrons) in the case of charmed particles. On top of the directed flow of the background medium an important contribution to the final signal arises in fact from the mismatch between the tilted geometry of the medium and the initial position of the $c\overline c$ pairs, symmetrically distributed around the beam axis. Predictions for the $D$-meson $v_1$ of the transport calculation of Ref.~\cite{Chatterjee:2017ahy} are shown in the right panel of Fig.~\ref{fig:v1}. The comparison with experimental data should allow one to probe the three-dimensional distribution of matter in heavy-ion collisions. Recently, some authors~\cite{Das:2016cwd,Chatterjee:2018lsx} have also proposed that the difference between the $v_1$ of $D^0$ and $\overline{D^0}$ mesons can be a unique probe of the huge electromagnetic fields present in the fireball during the deconfined phase; however, current experimental data~\cite{Adam:2019wnk} do not allow yet to draw firm conclusions.   

\section{In-medium hadronization and heavy-flavour hadrochemistry}
\begin{figure}[!ht]
\begin{center}
\includegraphics[clip,height=6cm]{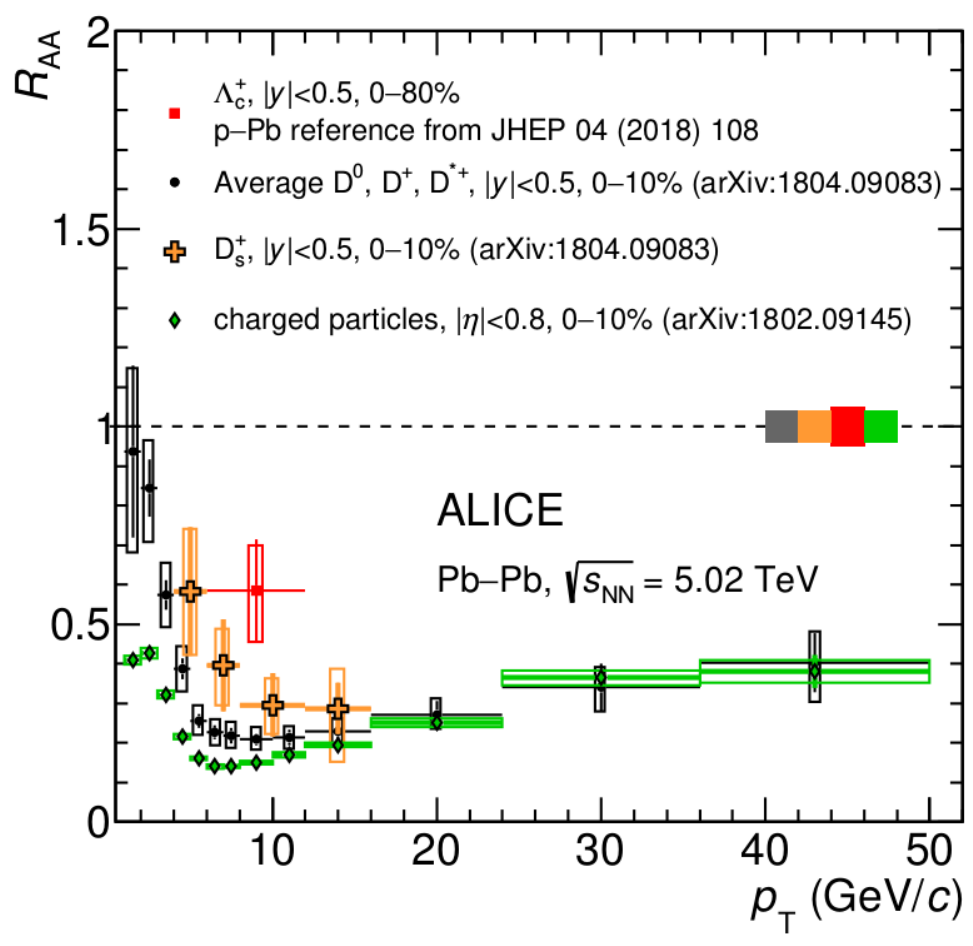}
\includegraphics[clip,height=6cm]{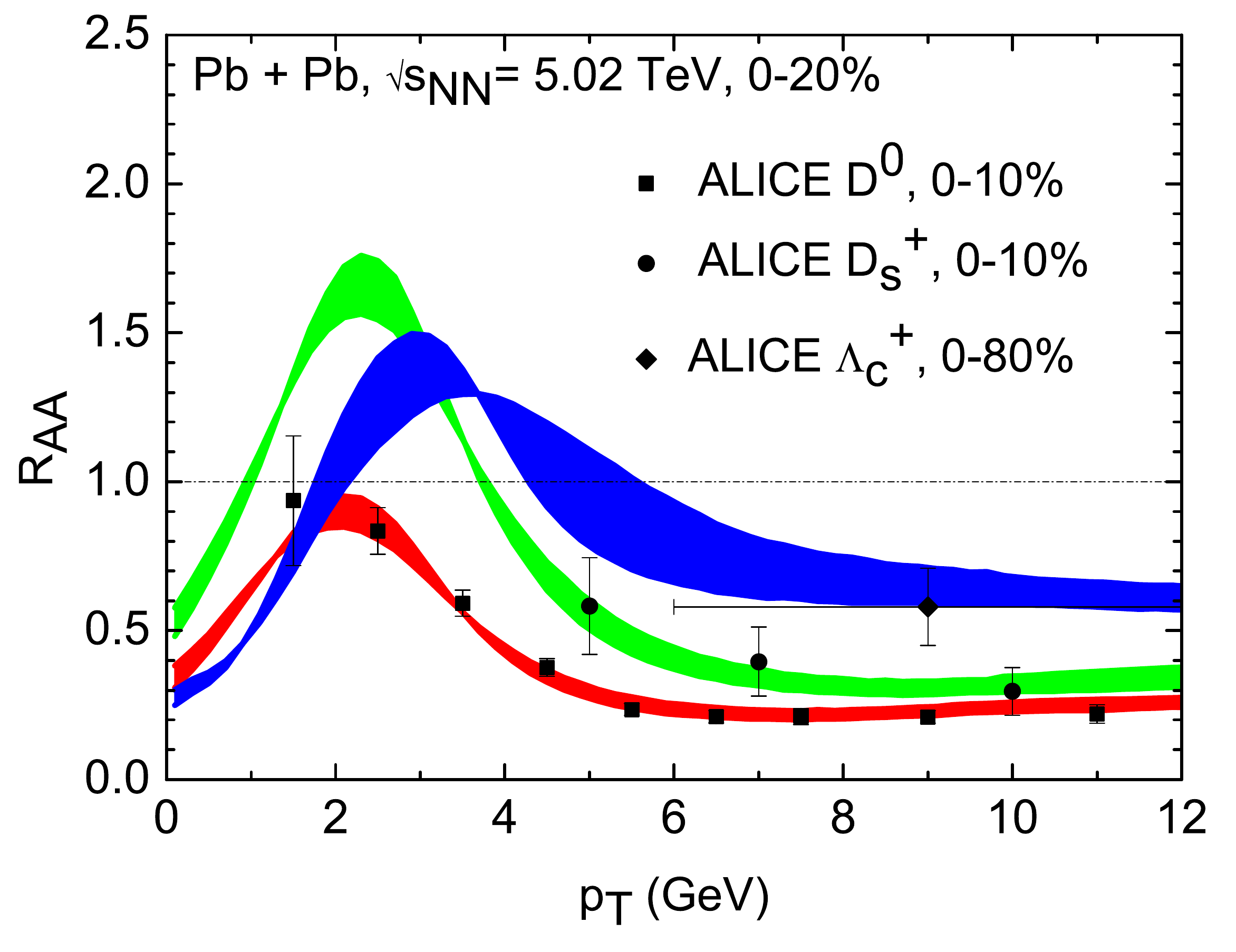}
\caption{Experimental results (left panel,~\cite{Acharya:2018ckj}) and transport-model predictions (right panel,~\cite{He:2019vgs}) for the nuclear modification factor of $D$ mesons, $D_s$ mesons and $\Lambda_c$ baryons in Pb-Pb collisions at the LHC.}\label{fig:hadrochemistry} 
\end{center}
\end{figure}
In elementary collisions (e.g. $e^++e^-,\;p+p...$) hadronization is described either in terms of vacuum fragmentation functions or, in the case of QCD event generators, through the decay of colour-singlet objects (clusters in Herwig, strings in PYTHIA). This mechanism in particular suppresses the production of strange hadrons and baryons, since one has to excite from the vacuum either $s\overline s$ pairs or diquark-antidiquark pairs, which are quite heavy. This holds also in the case of hadronization of charm or bottom quarks.

In heavy-ion collisions when the local temperature of the fireball is around the (de)confinement transition a heavy quark is surrounded by a lot of thermal partons. It is then conceivable that hadronization can occur differently than in the vacuum, through some form of recombination with the abundant (anti-)quarks (and possibly diquarks) nearby, with no penalty factor to produce a strange hadron or a baryon. In the case of charmed particles one may expect then a relative enhancement of the production of $D_s$ mesons and $\Lambda_c$ baryon with respect to proton-proton collisions. Various theoretical models were developed to account for in-medium hadronization and the consequent modification of heavy-flavour hadrochemistry, some based on a $N\to 1$ ($Q+\overline q\to M$ or $Q+qq\to B$) coalescence mechanism~\cite{Oh:2009zj,Plumari:2017ntm}, some others on the formation/decay of resonances (e.g. $Q+\overline q\leftrightarrow M$) in reactions parametrized by a Breit-Wigner cross-section~\cite{He:2019vgs}. In Fig.~\ref{fig:hadrochemistry} we show recent results obtained by the ALICE collaboration~\cite{Acharya:2018ckj}, in which the milder suppression of $D_s$ mesons and $\Lambda_c$ baryons compared to non-strange $D$-mesons suggests a relative enhancement of their yields compared to proton-proton collisions. In the right panel of Fig.~\ref{fig:hadrochemistry} we display recent predictions of one of the above theoretical calculations, in which the simulation of heavy quark transport throughout the fireball is interfaced to a hadronization mechanism based on resonance formation, leading to an enhanced production of $D_s$ and $\Lambda_c$~\cite{He:2019vgs}. 

\section{Conclusions and perspectives}
We tried to give an overview of the most recent achievements in the modelling of heavy-flavour production in heavy-ion collisions, emphasizing both the new theoretical developments and the application of transport calculations to the richer and richer set of observables which have become accessible in the last few years. A quite satisfactory description of most experimental results can be obtained, although for the final goal of getting a precise estimate of the heavy-quark transport coefficients we have still to wait for measurements of beauty at low $p_T$, where the non-relativistic dynamics allows a more solid comparison with theoretical results.  

\section*{References}
\bibliography{beraudo-proc}
\end{document}